\documentclass{article}
\pdfoutput=1
\usepackage{graphicx}
\usepackage{amsmath}
\usepackage{amssymb}
\usepackage{times}
\usepackage{url}
\usepackage{natbib}
\usepackage{chngpage}

\begin{document}
\title{Discovering general multidimensional associations}
\date{}
\maketitle

\begin{flushleft}

Ben Murrell$^{1,2,3\ast}$, 
Daniel Murrell$^{4}$, 
Hugh Murrell$^{5}$
\\
\bf{1} Computational Biology Group, Institute of Infectious Diseases and Molecular Medicine, University of Cape Town, Cape Town, South Africa
\\
\bf{2} Biomedical Informatics Research Division, eHealth Research and Innovation Platform, Medical Research Council, Cape Town, South Africa
\\
\bf{3} Department of Computer Science, Stellenbosch University, South Africa
\\
\bf{4} Unilever Centre for Molecular Sciences Informatics, Department of Chemistry, University of Cambridge, Lensfield Road, Cambridge, CB2 1EW, United Kingdom
\\
\bf{5} Department of Computer Science, University of KwaZulu-Natal, Pietermaritzburg, South Africa.
\\
$\ast$ E-mail: murrellb@gmail.com\\
\end{flushleft}

\textbf{When two variables are related by a known function, the coefficient of determination (denoted $R^2$) measures  the proportion of the total variance in the observations that is explained by that function. This quantifies the strength of the relationship between variables by describing what proportion of the variance is signal as opposed to noise. For linear relationships, this is equal to the square of the correlation coefficient, $\rho$. When the parametric form of the relationship is unknown, however, it is unclear how to estimate the proportion of explained variance equitably \citep{Reshef2011Detecting} - assigning similar values to equally noisy relationships. Here we demonstrate how to directly estimate a generalized $R^2$ when the form of the relationship is unknown, and we question the performance of the Maximal Information Coefficient (MIC) - a recently proposed information theoretic measure of dependence \citep{Reshef2011Detecting}. We show that our approach behaves equitably, has more power than MIC to detect association between variables, and converges faster with increasing sample size. Most importantly, our approach generalizes to higher dimensions, which allows us to estimate the strength of multivariate relationships ($Y$ against $A,B,\ldots$) and to measure association while controlling for covariates ($Y$ against $X$ controlling for $C$), whose importance was highlighted in \cite{Speed2011Correlation}.}\\



In \cite{Reshef2011Detecting}, the authors describe desired properties of a measure of bivariate association: generality and equitability. A measure that is general will discover, with sufficient sample size, any departure from independence, while a measure that is equitable will assign similar scores to equally noisy relationships of different kinds. A further attractive property is that a measure should scale like the coefficient of determination ($R^2$): the proportion of variance explained.

\cite{Reshef2011Detecting} demonstrates that other measures of association (including Spearman's rank correlation, mutual information, maximal correlation and principal curve-based correlation) are not equitable; different functional forms with similar amounts of noise can produce vastly different estimates of association strength. Here we show that generality and equability can be achieved by estimating a generalized $R^2$ through density approximation.

First consider a function with additive noise, $y = f(x) + \mathcal{N}$. The coefficient of determination is the proportion of variance in $y$ ``explained'' by the deterministic component $f(x)$ relative to the total variance in $y$, which is inflated by unexplained stochastic noise, $\mathcal{N}$. This notion of variance is defined in terms of average squared deviations - the distance between a data point and a model. The explained variance, $R^2$, is $1-\sigma^2_{Error}/\sigma^2_{Total}$, where
the total variance ($\sigma^2_{Total}$) is the average squared deviation from a flat ``null'' model and the error variance ($\sigma^2_{Error}$) is the average squared deviation from $f(x)$, the ``alternative'' model.


\begin{figure}[t]
\begin{center}
\begin{adjustwidth}[]{-0.75in}{-0.75in}
\includegraphics[width=6.5in]{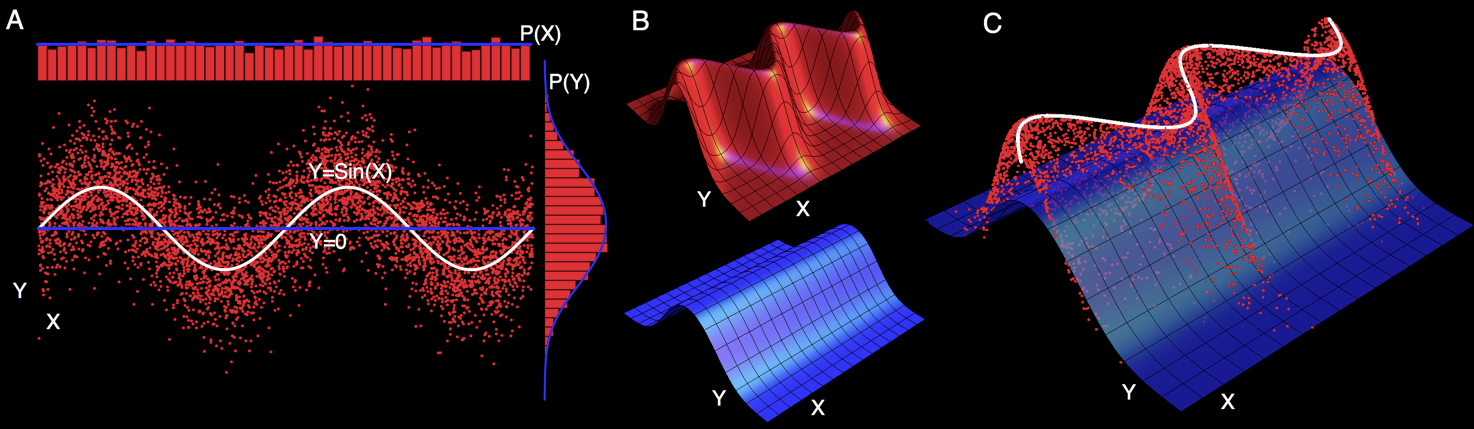}
\end{adjustwidth}
\end{center}
\caption{
{\bf Illustrating the generalized $\mathbf{R^2}$.} Panel {\textbf A}: Data is normally distributed about the alternative model - the  white regression line $Y=\sin(X)$. The null model is the blue $Y=0$. Marginal distributions of $X$ and $Y$ are represented above and to the right. The classical $R^2$ is calculated using deviations of the samples from the blue and white lines. Panel {\textbf B} depicts the probability distribution over $x_i,y_i$ for the alternative (red) and null (blue) models. Panel {\textbf C} shows the height of the observations on the alternative distribution, relative to the null distribution. The generalized $R^2$ is calculated from the ratio of these heights, and does not require an explicit regression line (white), which is included only as a guide for the eye.}
\label{Explain1}
\end{figure}

Least squares regression assumes that observations are normally distributed about the explanatory function. The deviation of a point from the regression line can thus be expressed as a probability density, and $R^2$ has an equivalent form \citep{Cox1989Analysis, Magee1990R,Nagelkerke1991Note}:

\begin{equation}
R^2 = 1- \prod_i  \left (\frac{P(x_i,y_i|null)}{P(x_i,y_i|alt)} \right ) ^{2/n}
\end{equation}
This formulation of $R^2$ asserts that the proportion of \emph{unexplained} variance is the geometric mean of the squared ratio of the probability of observing a data point under the null model over the probability of that data point under the alternative model. The explained variance is 1 minus the unexplained proportion. See figure 1 for a visual depiction.

Since this $R^2$ now depends only on the probability density ratio between two models, it is applicable even when the assumptions behind least squares regression are violated. This is a powerful rethinking of $R^2$. The idea of ``explained variance'' is generalized away from the restrictive assumptions of normally distributed noise, and, most importantly, the very notion of a regression curve is no longer required. This generalized $R^2$ can be calculated as long as the probability distributions for the null and alternative models can be evaluated.

We base our measure of dependence between variables upon this generalized $R^2$. Even when a known distribution generates our data, we still need to specify the null distribution before $R^2$ can be computed, but this generalized definition of $R^2$ is agnostic about a choice of null model. An attractive property for a measure of dependence is that it is 0 if and only if $X$ and $Y$ are independent. A sensible choice of null model is thus where $P(X,Y)=P(X)P(Y)$, enforcing independence. Since explicitly choosing a null distribution places a restriction on the generalized $R^2$, we distinguish our measure of association, calling it $A$. Classical $R^2$ from least squares regression assumes a different choice of null model (a constant function with normally distributed errors), so $A$ can be thought of as a sister to classical $R^2$. They are equivalent for bivariate Gaussian distributions, where the marginals are also normally distributed, but will differ (see SI1) when the null model for classical $R^2$ - a constant function with Gaussian errors - is a particularly bad fit. $A$ also has an information theoretic interpretation: for known distributions it is a sample estimate that converges to Linfoot's `Informational Measure of Correlation' \citep{Linfoot1957Informational} when the number of observations tends to infinity (see SI2).

\begin{figure}[t]
\begin{center}
\begin{adjustwidth}[]{-0.75in}{-0.75in}
\includegraphics[width=6in]{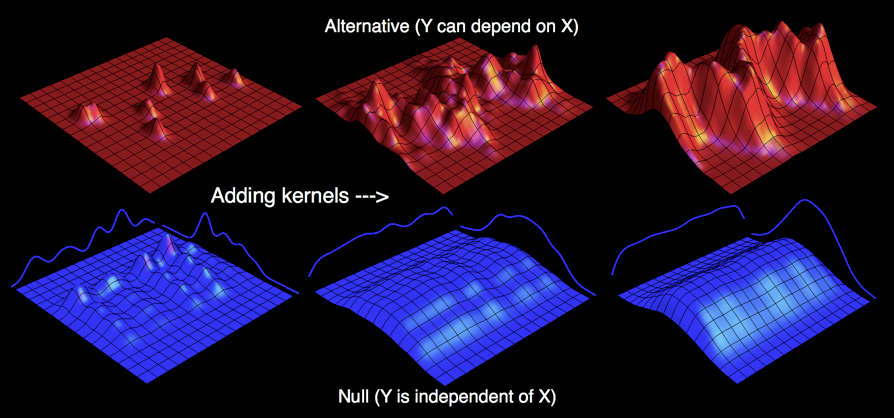}
\end{adjustwidth}
\end{center}
\caption{
{\bf Estimating an unknown distribution.} The distribution for the alternative model (red - where $X$ can depend on $Y$) is constructed by adding two dimensional Gaussian ``kernel'' distributions centered at each observation. As more of these kernels are added, the distribution comes to resemble the true distribution from which the observations are sampled. We can use a similar approach to estimating a null model that expressly disallows any dependence between $X$ and $Y$ (blue) by constructing one dimensional marginal distributions (the blue lines to either side) by summing one dimensional Gaussian kernels, and then creating the joint distribution as the product of these estimated marginals.}
\label{Explain2}
\end{figure}

\begin{figure}[!t]
\begin{center}
\begin{adjustwidth}[]{-0.75in}{-0.75in}
\includegraphics[width=6in]{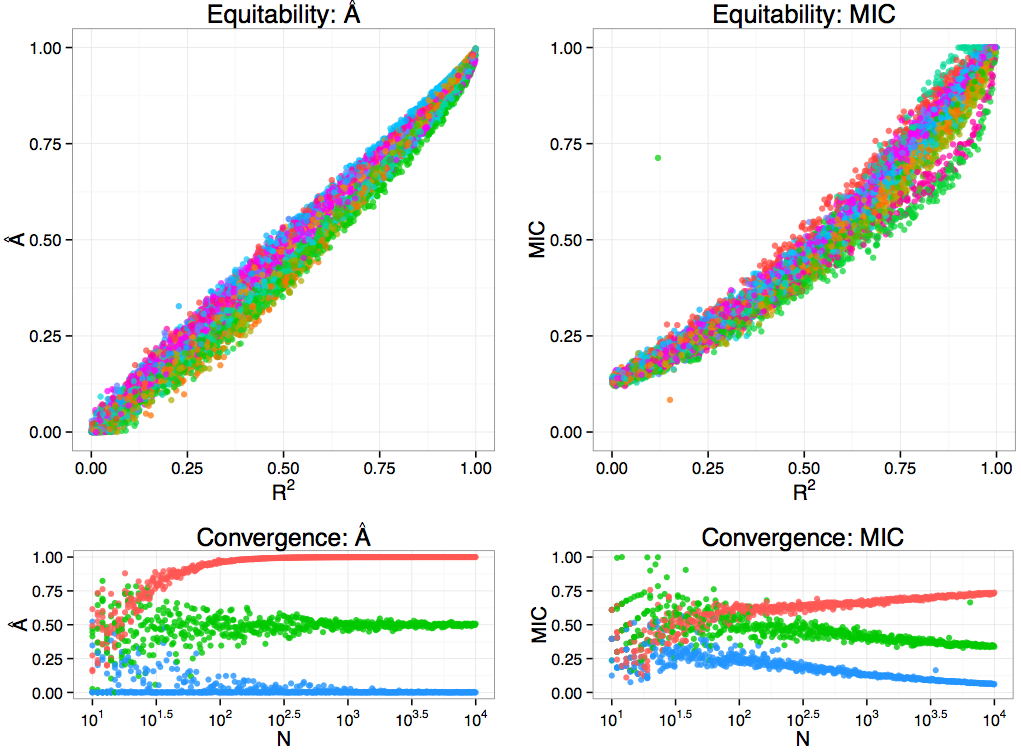}
\end{adjustwidth}
\end{center}
\caption{
{\bf Equitability and convergence of $\mathbf{\hat{A}}$}. Top: For functions of 2 variables, $\hat{A}$ is approximately equitable, as demonstrated with 16 example functions (see SI3 for a list). Each function is marked with a different color. $\hat{A}$ (left) is closer to the classical $R^2$ than MIC (right), especially for associations near independence. Bottom: Estimates of association from $\hat{A}$ (left) and MIC (right) as sample size ($N$) increases for three different relationships: a noiseless circle (red), a bivariate normal distribution with expected $R^2=0.5$ (green), and independent noise (blue). MIC converges very slowly.}
\label{equity2d}
\end{figure}

So far, the computation of $A$ requires a known distribution. Estimating $\hat{A} \approx A$ for a number of observations from an unknown distribution thus reduces to the problem of estimating the density at each point for an independent null and (potentially) dependent alternative model. We adopt a kernel density approach \citep{RosenblattM1956, Parzen1962Estimation}, where the density of the distribution at each point is approximated by the sum over a number of Gaussian `kernel' distributions centered at nearby points (see figure \ref{Explain2}). For the null model, we constrain the joint density to be the product of estimates of the marginal densities, enforcing independence. We wish to constrain  $\hat{A}$ to vary between 0 and 1, so we cannot allow the null to outperform the alternative model, lest  $\hat{A}$ become negative. We thus define the density of the alternative model at each sample point to be a weighted mixture of dependent (full joint) and independent (product of marginal) models, with a single mixture parameter controlling the proportion for all points. Thus the alternative model can reduce to the null model as a special case, ensuring non-negativity. We estimate the model parameters - and thus the densities - by maximizing the cross-validation likelihood (see Methods).

\begin{figure}[t]
\begin{center}
\begin{adjustwidth}[]{-0.75in}{-0.75in}
\includegraphics[width=6in]{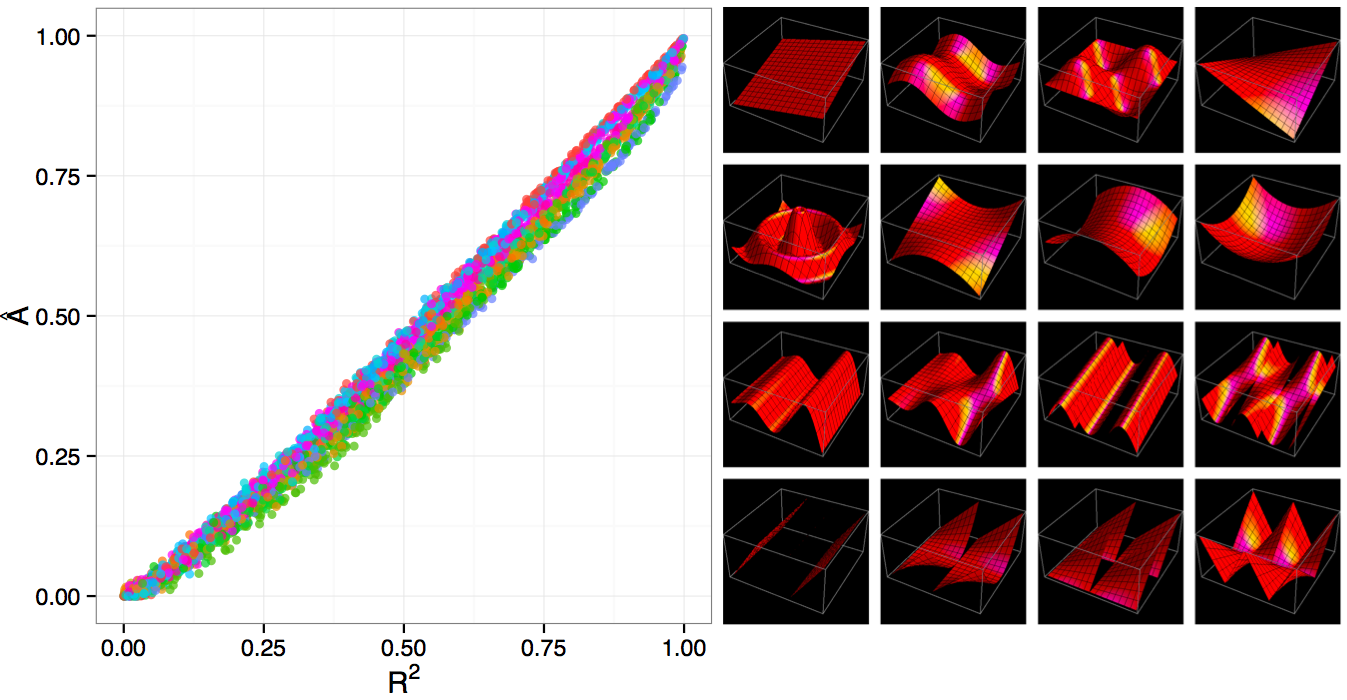}
\end{adjustwidth}
\end{center}
\caption{
{\bf Equitability of $\mathbf{\hat{A}}$ in higher dimensions}. $\hat{A}$ against $R^2$ (left) for multivariate datasets generated by adding normally distributed noise to 16 different functions of two variables (right - see SI3 for more detail).}
\label{equity3D}
\end{figure}

Figure \ref{equity2d} demonstrates that $\hat{A}$ is approximately equitable across a number of relationships (see SI3 for details), and is in greater agreement with classical $R^2$ than is MIC, especially for relationships where $R^2$ is close to 0. When the marginal distribution of a variable departs substantially from a normal distribution, $\hat{A}$ (like MIC) may produce more conservative estimates of association than classical $R^2$ (see SI1). This is because the null model for $\hat{A}$ makes less restrictive assumptions (only independence is assumed, without a parametric form), describing the data better than the null model for classical $R^2$, which is a constant function with Gaussian errors.

$\hat{A}$ converges faster with increasing sample size than MIC (figure \ref{equity2d} - bottom panels). For example, despite having a theoretical large-sample limit of 1 for a noiseless circle \citep{Reshef2011Detecting}, $MIC \approx 0.74$ when $N=10000$. In contrast, $\hat{A} \gtrsim 0.99$ when $N \ge 200$.

We also introduce a statistical test for non-independence associated with $\hat{A}$, computing $p$-values through a randomization procedure (See SI4 for details). The $\hat{A}$ test has greater power to detect associations than MIC for all but one of the relationships we tested, and outperforms Sz\'{e}kely's dCov test for association \citep{Szekely2009Brownian} for all non-linear relationships tested. The $\hat{A}$ test was more comparable in performance to the recently proposed HHG test \citep{Heller2012Consistent}, having greater power on 4 out of 7 tested relationships. See SI4 for further results.

As shown in figure \ref{equity3D}, $\hat{A}$ generalizes to multiple dimensions, producing equitable estimates very similar to classical $R^2$ for functions of two dimensions. It can assess the strength of association between vector valued variables, indicating what proportion of the variance in $(X,Y)$ is explained by $(A,B,C)$, for example. It also generalizes to more than two variables (with each variable being possibly vector valued), which could be used to discover lower dimensional manifolds embedded in a higher dimensional space (see SI5).

As pointed out in \cite{Speed2011Correlation}, an important question is how much of the variance in $Y$ can be explained by $X$, after controlling for $C$. Here we introduce a non-linear analog of the semipartial correlation coefficient. We show (see SI6) that this agrees with the linear semipartial correlation when all relationships are linear. When the relationships are non-linear, however, the standard linear semipartial correlation can severely underestimate semipartial association, but it can also overestimate the semipartial association between $Y$ and $X$ by ignoring a non-linear dependence of $Y$ on the control variable $C$, returning values close to 1 when in fact $Y$ is conditionally independent of $X$ given $C$. Our non-linear semipartial association has no such difficulty, returning values close to 0 for such cases.

While this paper represents the initial practical contribution, further work remains to characterize the theoretical properties of $A$ and $\hat{A}$. $A$ is clearly invariant to monotonic transformations of variables, but its estimate $\hat{A}$ is not, although it may be as $N$ tends to infinity. Simulations suggest that $\hat{A}$ tends to 0 wherever variables are independent, and 1 whenever a relation is noiseless and nowhere flat, but perhaps there are other circumstances under which 1 will be the large sample limit (MIC, for example, can achieve 1 at large samples for noisy relationships - see SI7). Is the $\hat{A}$ test for independence consistent against all alternatives, achieving a power of $100\%$ as $N$ tends to infinity whenever independence is violated in any way? $\hat{A}$ appears to be robust to outliers (see SI8), but is it possible to design outlier distributions that mislead it? $\hat{A}$ could also be improved by more sophisticated techniques to estimate the density ratio of the joint and independent distributions \citep{Suzuki2009Mutual, Vincent02manifoldparzen}, which may improve the convergence of $\hat{A}$ for smaller sample sizes, but at a computational cost.

\section*{Methods}
Consider two (possibly) vector valued variables, $\mathbf{X}$ and $\mathbf{Y}$, with $n$ observations $\{\mathbf{x}_1,\ldots,\mathbf{x}_n\}$ and $\{\mathbf{y}_1,\ldots,\mathbf{y}_n\}$. Each $\mathbf{x}_i$ itself may be a vector $x^a_i,\ldots,x^z_i$, as may each $\mathbf{y}_i$. Further, imagine three kernel distributions, $K_X(\mathbf{x})$, $K_Y(\mathbf{y})$ and $K_{XY}(<\mathbf{x,y}>)$, where the kernels are symmetric, non-negative, and integrate to 1, and where angle brackets indicate vector concatenation. Our null model assumes that  $\mathbf{X}$ and $\mathbf{Y}$ are independent, and so we define the leave-one-out cross validation likelihood as the product of marginal kernel density estimates:

\begin{eqnarray*}
L_{CV}(null)&=&\prod_{i=1}^n P(\mathbf{x}_i |\mathbf{x}_{\forall j \ne i}) P(\mathbf{y}_i |\mathbf{y}_{\forall j \ne i})\\
            &\approx&\prod_{i=1}^n \left [ \sum_{\forall j \ne i} \frac{K_X(\mathbf{x}_j-\mathbf{x}_i)}{n-1} \times \sum_{\forall j \ne i} \frac{K_Y(\mathbf{y}_j-\mathbf{y}_i)}{n-1} \right ]
\end{eqnarray*}

The alternative model allows $\mathbf{Y}$ to depend on $\mathbf{X}$ for a proportion of points, $w$, with a leave-one-out cross validation likelihood defined as:
\begin{adjustwidth}[]{-0.75in}{-0.75in}
\begin{eqnarray*}
L_{CV}(alt)&=&\prod_{i=1}^n \left[ w \times P(\mathbf{x}_i,\mathbf{y}_i |\mathbf{x}_{\forall j \ne i},\mathbf{y}_{\forall j \ne i})+ (1-w)P(\mathbf{x}_i |\mathbf{x}_{\forall j \ne i}) P(\mathbf{y}_i|\mathbf{y}_{\forall j \ne i}) \right]\\
            &\approx&\prod_{i=1}^n   \left [ w  \sum_{\forall j \ne i} \frac{K_{XY}(<\mathbf{x}_j-\mathbf{x}_i,\mathbf{y}_j-\mathbf{y}_i>)}{n-1} + (1-w) \ \sum_{\forall j \ne i} \frac{K_X(\mathbf{x}_j-\mathbf{x}_i)}{n-1} \sum_{\forall j \ne i} \frac{K_Y(\mathbf{y}_j-\mathbf{y}_i)}{n-1}  \right]
\end{eqnarray*}
\end{adjustwidth}

In our particular implementation, the values of each variable are replaced with their ranks (this is for computational convenience and should have little effect since $A$ itself is invariant to order preserving transformations of variables), and the kernels are isotropic Gaussians, with $K_X$ and $K_Y$ sharing an `independent' kernel variance parameter $\sigma^2_I$, and $K_{XY}$ having a distinct `dependent' variance parameter, $\sigma^2_D$. The null model thus has a single parameter, $\sigma^2_I$, and the alternative model has 3 parameters: $\sigma^2_I$, $\sigma^2_D$, and $w$. These parameters are optimized numerically to maximize the cross-validation likelihood, yielding $\hat{A}$ after employing equation 1. We found this estimate to be slightly biased down (relative to classical $R^2$ for bivariate Gaussians, where $A$=classical $R^2$), especially for small samples, so we included an empirically-estimated small samples correction (see SI9). An R package named ``matie'' (Measuring Association and Testing Independence Efficiently" - see SI10) for estimating $\hat{A}$ is available on CRAN (\url{http://cran.r-project.org/web/packages/matie/}). Like MIC, estimating $\hat{A}$ is quadratic in the sample size, but with a much lower growth rate than MIC (see SI11). Supporting Information can be found at: \url{http://www.cs.sun.ac.za/~bmurrell/Murrell_Matie_SI.pdf}.


\bibliographystyle{apalike}
\bibliography{Maatie}

\begin{thebibliography}{}

\bibitem[Cox and Snell, 1989]{Cox1989Analysis}
Cox, D.~R. and Snell, E.~J. (1989).
\newblock {\em {Analysis of Binary Data, Second Edition (Chapman \& Hall/CRC
  Monographs on Statistics \& Applied Probability)}}.
\newblock Chapman and Hall/CRC, 2 edition.

\bibitem[Heller et~al., 2012]{Heller2012Consistent}
Heller, R., Heller, Y., and Gorfine, M. (2012).
\newblock {A consistent multivariate test of association based on ranks of
  distances}.
\newblock {\em Biometrika}.

\bibitem[Linfoot, 1957]{Linfoot1957Informational}
Linfoot, E.~H. (1957).
\newblock {An informational measure of correlation}.
\newblock {\em Information and Control}, 1(1):85--89.

\bibitem[Magee, 1990]{Magee1990R}
Magee, L. (1990).
\newblock {R 2 Measures Based on Wald and Likelihood Ratio Joint Significance
  Tests}.
\newblock {\em The American Statistician}, 44(3):250--253.

\bibitem[Nagelkerke, 1991]{Nagelkerke1991Note}
Nagelkerke, N. J.~D. (1991).
\newblock {A note on a general definition of the coefficient of determination}.
\newblock {\em Biometrika}, 78(3):691--692.

\bibitem[Parzen, 1962]{Parzen1962Estimation}
Parzen, E. (1962).
\newblock {On Estimation of a Probability Density Function and Mode}.
\newblock {\em The Annals of Mathematical Statistics}, 33(3):1065--1076.

\bibitem[Reshef et~al., 2011]{Reshef2011Detecting}
Reshef, D.~N., Reshef, Y.~A., Finucane, H.~K., Grossman, S.~R., McVean, G.,
  Turnbaugh, P.~J., Lander, E.~S., Mitzenmacher, M., and Sabeti, P.~C. (2011).
\newblock {Detecting Novel Associations in Large Data Sets}.
\newblock {\em Science}, 334(6062):1518--1524.

\bibitem[Rosenblatt, 1956]{RosenblattM1956}
Rosenblatt, M. (1956).
\newblock {Remarks on Some Nonparametric Estimates of a Density Function}.
\newblock {\em The Annals of Mathematical Statistics}, 27(3):832--837.

\bibitem[Speed, 2011]{Speed2011Correlation}
Speed, T. (2011).
\newblock {A Correlation for the 21st Century}.
\newblock {\em Science}, 334(6062):1502--1503.

\bibitem[Suzuki et~al., 2009]{Suzuki2009Mutual}
Suzuki, T., Sugiyama, M., and Tanaka, T. (2009).
\newblock {Mutual information approximation via maximum likelihood estimation
  of density ratio}.
\newblock In {\em 2009 IEEE International Symposium on Information Theory},
  pages 463--467. IEEE.

\bibitem[Sz\'{e}kely and Rizzo, 2009]{Szekely2009Brownian}
Sz\'{e}kely, G.~J. and Rizzo, M.~L. (2009).
\newblock {Brownian distance covariance}.
\newblock {\em The Annals of Applied Statistics}, 3(4):1236--1265.

\bibitem[Vincent and Bengio, 2002]{Vincent02manifoldparzen}
Vincent, P. and Bengio, Y. (2002).
\newblock Manifold parzen windows.
\newblock In {\em Advances in Neural Information Processing Systems 15}, pages
  825--832. MIT Press.

\end{thebibliography}



\end{document}